\pdfoutput=1
\documentclass[aps,prl,superscriptaddress,showpacs,floatfix,twocolumn]{revtex4-1}
\usepackage{graphicx,amsmath,amssymb,xspace,epsfig,float,multirow,subfigure,tabularx}

\begin{document}

\title{Strong thermal fluctuations in cuprate superconductors in magnetic
field above $T_{c}$}
\author{Xujiang Jiang}
\affiliation{School of Physics, Peking University, Beijing
100871, \textit{China}.} 
\author{Dingping Li}
\email{lidp@pku.edu.cn}
\affiliation{School of Physics, Peking University, Beijing
100871, \textit{China}.} 
\author{Baruch Rosenstein}
\email{vortexbar@yahoo.com }
\affiliation{Electrophysics
Department, National Chiao Tung University, Hsinchu 30050, \textit{Taiwan,
R. O. C}.}
\date{\today}

\begin{abstract}
Recent measurements of fluctuation diamagnetism in high temperature
superconductors show distinct features above and below $T_{c}$, which can
not be explained by simple gaussian fluctuation theory. Self consistent
calculation of magnetization in layered high temperature superconductors,
based on the Ginzburg-Landau-Lawrence-Doniach model and including all Landau
levels is presented. The results agree well with the experimental data in
wide region around $T_{c}$, including both the vortex liquid below $T_{c}$
and the normal state above $T_{c}$. The gaussian fluctuation theory
significantly over-estimates the diamagnetism for strong fluctuations. It is
demonstrated that the intersection point of magnetization curves appears in
the region where the lowest Landau level contribution dominates.
\end{abstract}

\maketitle

\textit{Introduction.} One of the numerous qualitative differences between
high $T_{c}$ superconductors (HTSC) and low $T_{c}$ superconductors
is dramatic enhancement of thermal fluctuation effects. The thermal
fluctuations are much stronger in HTSC not just due to higher critical
temperatures, much shorter coherence length and high anisotropy play a
major role in the enhancement too. Since thermal fluctuations are strong the
effect of superconducting correlations (pairing) can extend into the normal
state well above the critical temperature. The normal state properties of
the underdoped cuprates exhibit a number of anomalies collectively referred
to as the "pseudogap" physics \cite{Timusk99} and their physical origin is
still poorly understood. It is natural therefore to attempt to associate
some of these phenomena with the superconducting thermal fluctuations or
"preformed" Cooper pairs\cite{Emery95}.

The interest in fluctuations was invigorated after the Nernst effect was
observed \cite{Xu00} all the way up to the pseudogap crossover temperature $%
T^{\ast }$ in underdoped $La_{2-x}Sr_{x}CuO_{4}$ ($LSCO$). Assuming
that Nernst effect is primarily due to thermal fluctuations, the whole
pseudogap region would be associated with preformed Cooper pairs and
become a precursor of the superconducting state. The finding motivated
additional experiments on Nernst effect in various HTSC \cite{Wang06}, as
well as renewed study of thermal fluctuations in the temperature region
between $T_{c}$ and $T^{\ast }$ by other probes: electric \cite{Rullier11} and
thermal conductivity \cite{Kudo04} and diamagnetism \cite{LuLi10}. The
main goal was to try to quantify the superconducting fluctuation effects, so
they can be either directly linked or separated from the pseudogap physics.
This requires a reliable quantitative theory of influence of thermal
fluctuations on transport (Nernst effect, thermal and electric conductivity)
and thermodynamic (magnetization, specific heat) physical quantities. Since
there is no sufficiently simple or/and widely accepted microscopic theory of
HTSC, one has to rely on a more phenomenological Ginzburg - Landau (GL)
theory \cite{Larkin05} that, although not sensitive to microscopic details,
is accurate and simple enough to describe the fluctuations above $T_{c}$.
While the transport experiments like Nernst effect have some hotly debated
experimental \cite{Chang12} and theoretical \cite{Varlamov09} issues to be
addressed, the clearest data come from recent thermodynamical measurements
of magnetization \cite{Kivelson10} in $LSCO$, $Bi_{2}Sr_{2}CaCu_{2}O_{8+%
\delta }$ ($BSCCO$) and $YBa_{2}Cu_{3}O_{7}$ ($YBCO$) \cite{LuLi10}.

\begin{figure}[tbp]
\centering
\includegraphics[width=8cm]{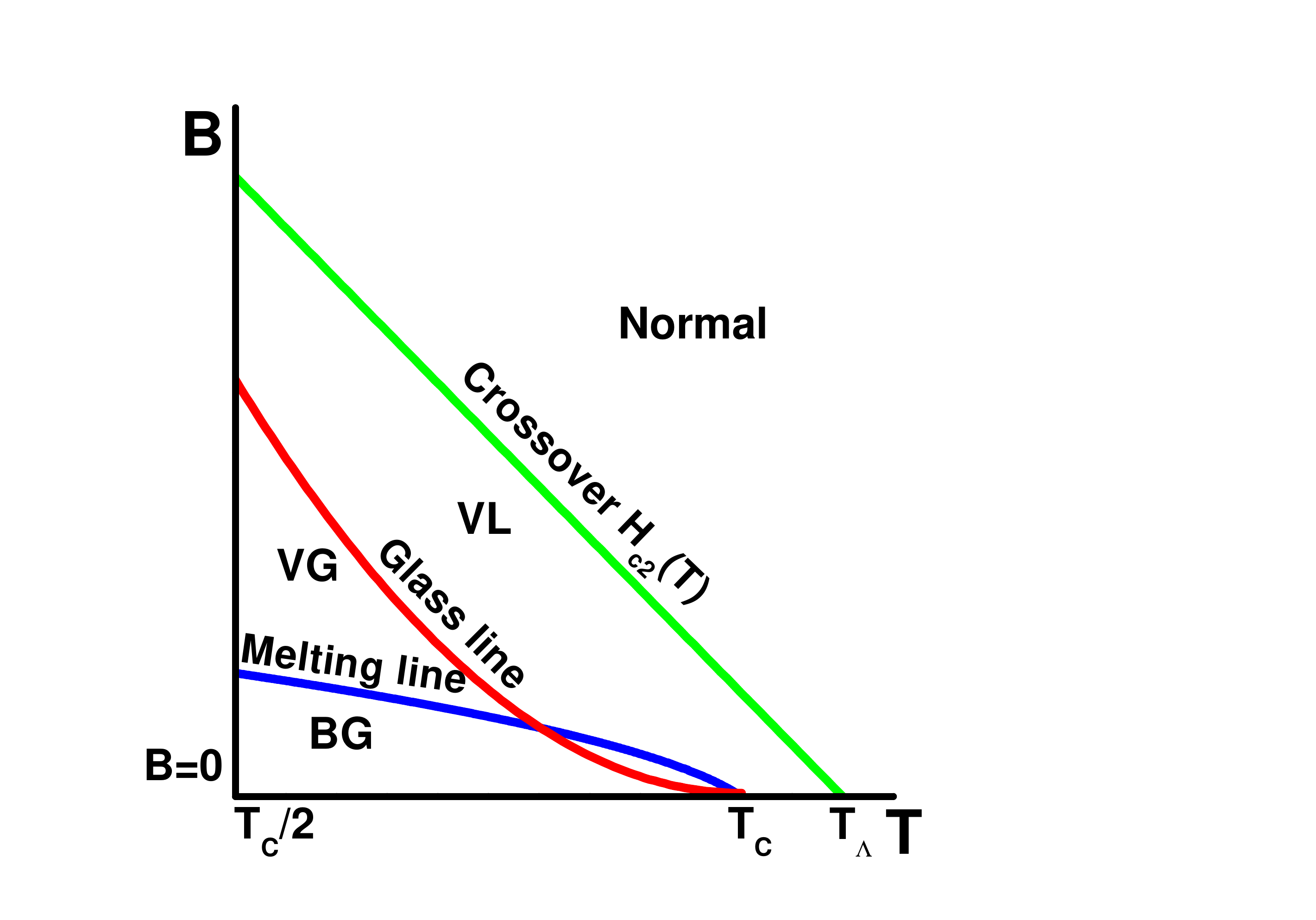}
\vspace{-0.5cm}
\caption{Magnetic phase diagram of high $T_{c}$ superconductors. VL is the
vortex liquid region, while VG and BG are the vortex glass and Bragg glass. }
\end{figure}

The purpose of this note is to provide a convincing theoretical description
of the magnetization data. Our conclusion is that the GL description of the
layered materials $LSCO$, $BSCCO$ and $YBCO$ by the Lawrence - Doniach model
within the self consistent fluctuation theory (SCFT, sometimes refered to
as Hartree approximation) fits well the fluctuation effects in major
families of HTSC materials in wide range of fields and temperatures and
demonstrates that the fluctuation effects extend to well above $T_{c}$ far
below $T^{\ast }$. This means that there is no evidence that the pseudogap
physics influences the diamagnetism and that superconductivity probably
plays no role at $T^{\ast }$.

Strong diamagnetism of a type II superconductor takes a form of network of
Abrikosov flux lines (vortices) created by magnetic field. Vortices strongly
interact with each other creating highly correlated configurations. A
generic magnetic phase diagram of HTSC \cite{Li03}, Fig.1, contains four
phases: two inhomogeneous phases, unpinned crystal and pinned Bragg glass
and two homogeneous phases, unpinned vortex liquid and pinned vortex glass.
In HTCS thermal fluctuations are strong enough to melt the lattices \cite%
{Zeldov95} into the vortex liquid over very large portion of the phase
diagram. This portion covers the fields and temperatures of the above
experiments, both above and below $T_{c}$ and for fields up to $40T$. The
glass line separates pinned vortex matter (zero resistivity) from the
unpinned one (nonzero resistivity due to flux flow).

Fluctuation diamagnetism in type II superconductors has been studied
theoretically\cite{Larkin05} within both the microscopic theory (starting
from the pioneering work of Aslamazov and Larkin) and the GL approach. In
all of these calculations (with an exception of the strong field limit that
allows the lowest Landau level approximation, see \cite{Li10}) the
fluctuations were assumed to be small enough, so they can be taken into
account perturbatively. Within the GL approach this is referred to as the
gaussian fluctuation theory (GFT)\cite{Larkin05,Carballeira00}. The GFT
applied to the recent HTSC magnetization data was criticized \cite{Ong07} to
fit just a single curve (magnetic field) rather than a significant portion
of the magnetic phase diagram near $T_{c}$. To determine theoretically
fluctuation diamagnetism for strong thermal fluctuations, one therefore must
go beyond this simple approximation neglecting the effect of the quartic
term in the GL free energy. The effect of the quartic term is taken into
account within SCFT, widely used in physics of phase transitions at zero
magnetic field and was adapted to transport property in magnetic field
\cite{Dorsey91,Tinh09}. Since disorder is not considered, our results
are limited to the vortex liquid phase of the magnetic phase diagram of
Fig.1, where vortices are depinned.

\textit{The GL model of layered superconductor.} Layered superconductor is
sufficiently accurately described on the mesoscopic scale by the
Lawrence-Doniach free energy (incorporating microscopic thermal fluctuation
via dependence of parameters on temperature $T$, but not containing thermal
fluctuations of the order parameter on the mesoscopic scale):
\begin{eqnarray}
F\left[ \psi \right]  &=&s^{\prime }\sum_{l}\int_{\mathbf{r}}\left[ \frac{%
\hbar ^{2}}{2m_{a}}\left\vert \mathbf{D}\psi _{l}\right\vert ^{2}+\frac{%
\hbar ^{2}}{2m_{c}d^{\prime 2}}\left\vert \psi _{l}-\psi _{l+1}\right\vert
^{2}\right.   \notag \\
&&\left. +\alpha \left( T-T_{\Lambda }\right) \left\vert \psi
_{l}\right\vert ^{2}+\frac{\beta }{2}\left\vert \psi _{l}\right\vert ^{4}%
\right] \text{.}
\end{eqnarray}%
Here $\psi _{l}\left( x,y\right) $ is the order parameter in the $l^{\text{th%
}}$ layer, $\mathbf{D\equiv \triangledown }+\frac{ie^{\ast }}{\hbar c}%
\mathbf{A}$, is the covariant derivative ($e^{\ast }=2|e|$) and $\mathbf{A}$
is the vector potential of magnetic field oriented along the
crystallographic $c$ axis. The (effective) layer thickness is $s^{\prime }$
and the distance between the layers - $d^{\prime }$. Note that the
temperature $T_{\Lambda }$, that will be called "mean field" or "bare"
transition temperature, is larger than the real transition temperature $T_{c}
$.

\begin{figure*}[tph]
\begin{center}
\subfigure[LSCO]{\includegraphics[width=7cm]{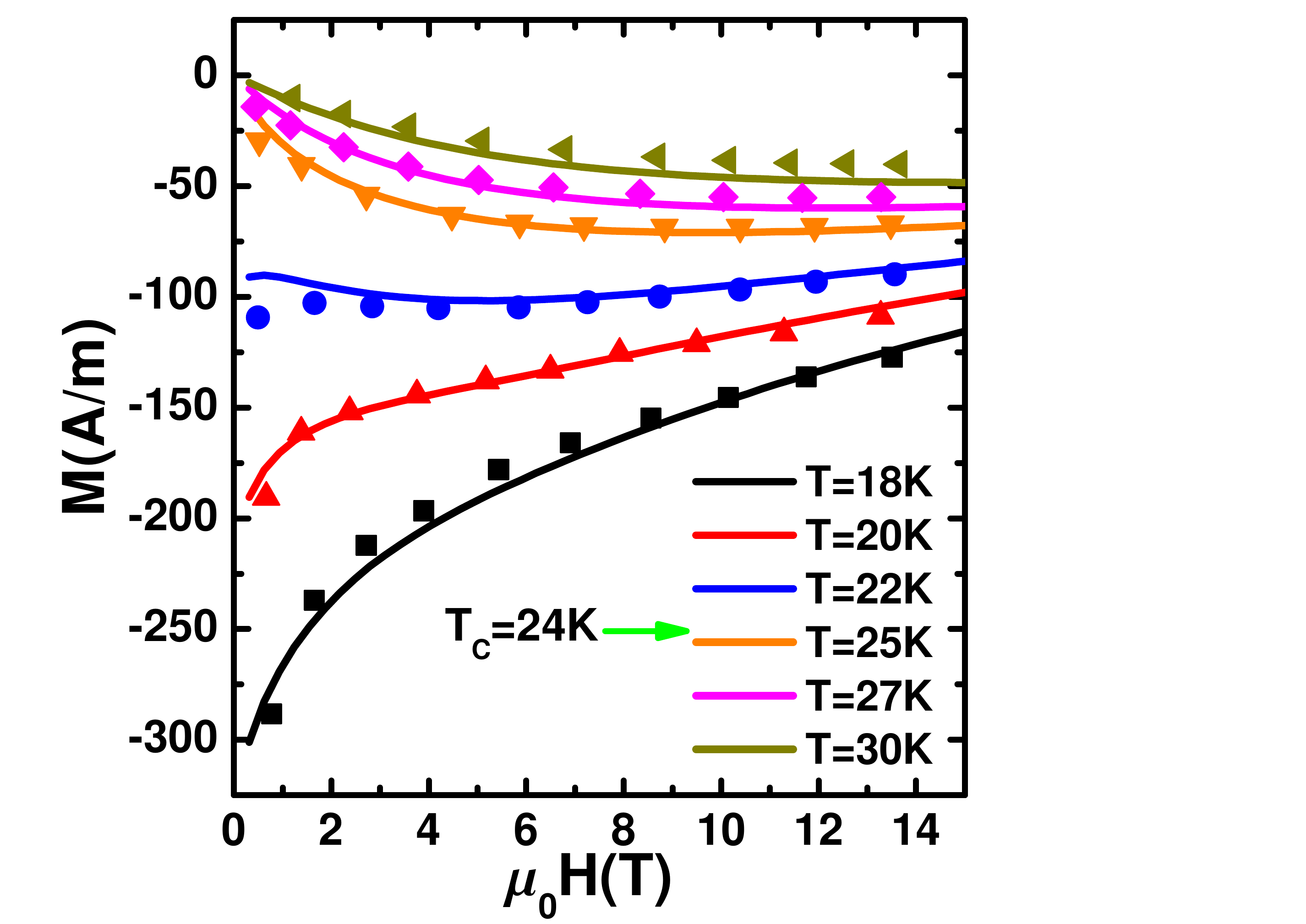}} \hspace{-1.8cm} %
\subfigure[BSCCO]{\includegraphics[width=7cm]{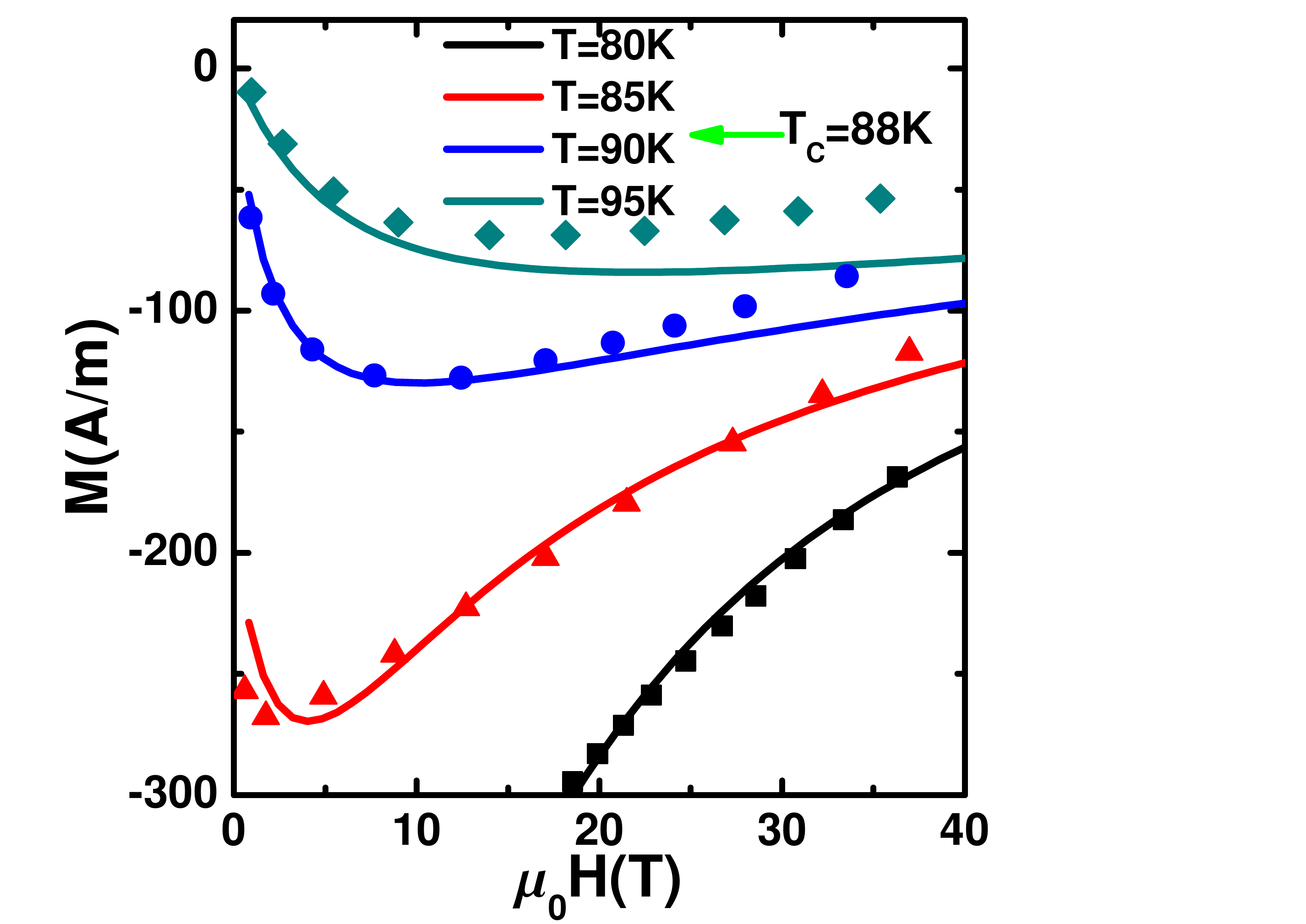}} \hspace{-1.8cm} %
\subfigure[YBCO]{\includegraphics[width=7cm]{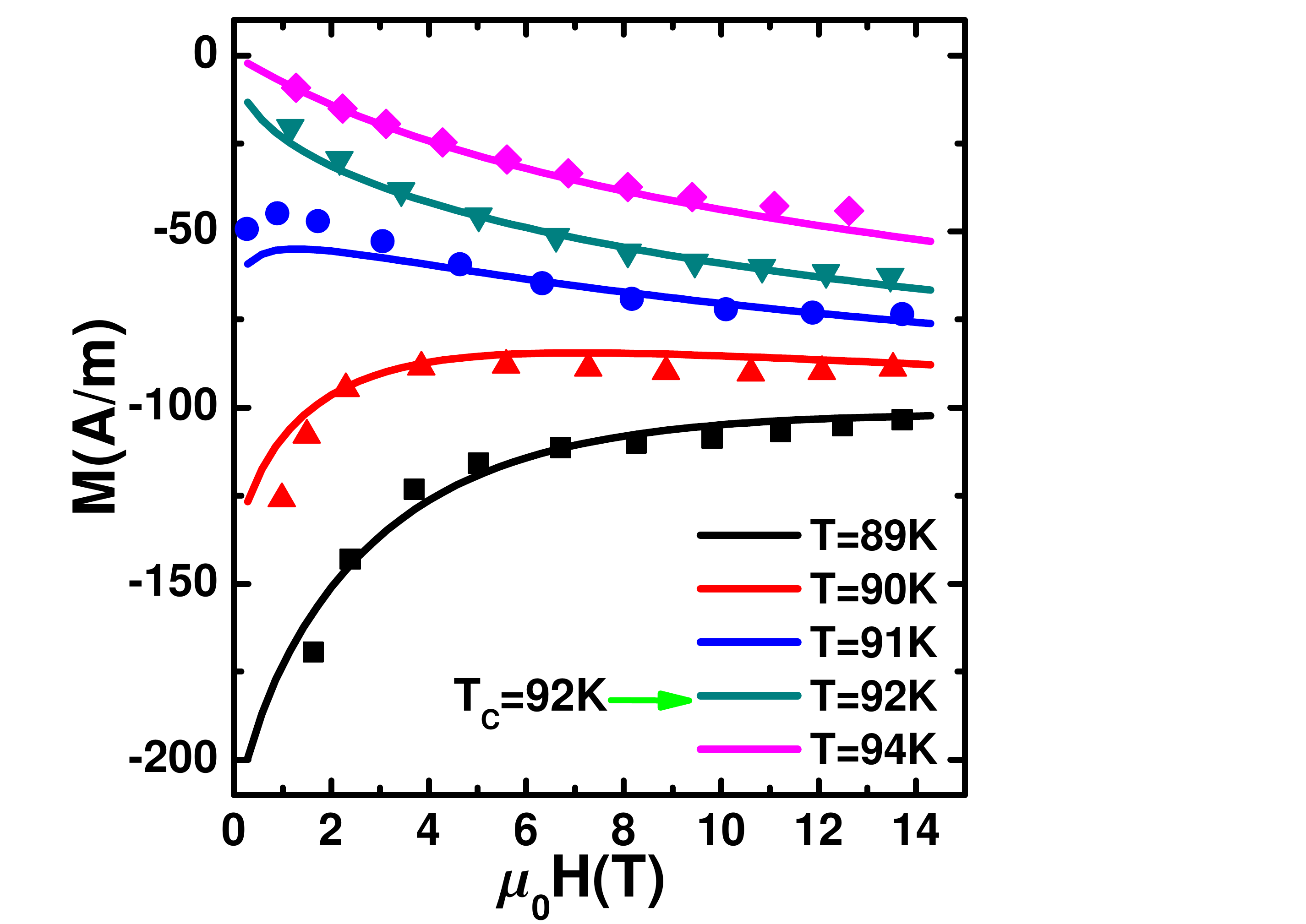}}
\end{center}
\par
\vspace{-0.5cm}
\caption{ Magnetization data of ref.\protect\cite{LuLi10} (dots) and their
self consistent approximation fits (solid lines). Three major families of
high $T_{c}$ superconductors are represented: (a) underdoped LSCO, (b)
optimally doped BSCCO, (c) optimally doped YBCO. The curve closest to $T_{c}$
for each sample were used to determine the fitting parameters given in Table
I. Each set of curves uses just three fitting parameters.}
\end{figure*}
\begin{table*}[tph]
\caption{Fitting parameters for LSCO, BSCCO, and YBCO.}
\begin{center}
\renewcommand\arraystretch{1.5}
\begin{tabularx}{\textwidth}{XXXXXXXXX}
\hline
\hline
$Material$ & $T_c(Kelvin)$ & $d^{\prime}({Angstrom})$ &$H_{c2}(Tesla)$ & $T_{\Lambda}(Kelvin)$ & $\gamma$ & $\Lambda$ &$\kappa$& $Gi$ \\
\hline
$LSCO$ & $24$ & $6.58$ &$31$ & $33$ & $29$ & $0.30$ & 66.7 &$0.033$ \\
$BSCCO$ & $88$ & $19.6$ &$115$ & $99$ & $19$ & $0.25$ & 55.6 &$0.025$ \\
$YBCO$ & $92$ & $11.68$ &$220$ & $100$ & $4.1$ & $0.22$ & 78.7 &$0.0026$ \\
\hline
\hline
\end{tabularx}
\end{center}
\end{table*}

The "bare" coherence length $\xi =\hbar /\sqrt{2m_{a}\alpha T_{\Lambda }}$
will be used as the unit of length and the upper critical field $%
H_{c2}\equiv \hbar c/e^{\ast }\xi ^{2}$ as the magnetic field unit. They
depend on coarse graining scale (cutoff scale $\Lambda $) at which the
mesoscopic model is derived (in principle) from The dimensionless order
parameter is $\phi =\sqrt{\beta /2\alpha T_{\Lambda }}\psi $, so that the GL
Boltzmann factor in scaled units takes a form,
\begin{eqnarray}
f &=&\frac{F}{T}=\frac{1}{2\omega _{\Lambda }t_{\Lambda }}\sum_{l}\int_{%
\mathbf{r}}\left[ \left\vert \mathbf{D}\phi _{l}\right\vert
^{2}+d^{-2}\left\vert \phi _{l}-\phi _{l+1}\right\vert ^{2}\right.   \notag
\\
&&\left. -\left( 1-t_{\Lambda }\right) \left\vert \phi _{l}\right\vert
^{2}+\left\vert \phi _{l}\right\vert ^{4}\right] \text{.}
\end{eqnarray}%
Here $t_{\Lambda }=T/T_{\Lambda }$, $b=B/H_{c2}$ are dimensionless
temperature and induction.It is more convenient to use the fluctuation
strength parameter $\omega _{\Lambda }=\sqrt{2Gi_{\Lambda }}\pi /s$, instead
of the more customary ("bare") Ginzburg number $Gi_{\Lambda }=2\left(
e^{\ast }/c\hbar \right) ^{3}\kappa ^{4}T_{\Lambda }^{2}\gamma ^{2}/H_{c2}$.
Since the renormalization by strong thermal fluctuations is central in this
work, bare quantities carry index $\Lambda $, although the results used for
fitting experiments will utilize renormalized parameters. The anisotropy $%
\gamma =\sqrt{m_{c}/m_{a}}$, $s=s^{\prime }\gamma /\xi _{\Lambda }$ and $%
d=d^{\prime }\gamma /\xi _{\Lambda }$. In strongly type II suprconductors
the Ginzburg parameter $\kappa =\lambda /\xi >>1$ and magnetic field is
nearly homogeneous\cite{suppl}, so we choose the Landau gauge $\mathbf{A}%
=(-by,0)$ in $\mathbf{D}=\mathbf{\nabla }+i\mathbf{A}$.

\textit{Fluctuation diamagnetism calculated within SCFT.} The idea the
method \cite{Li10} is as follows\cite{suppl}. Let us divide the GL Boltzmann
factor $f\left[ \phi \right] $ into an optimized quadratic ("large") part,
\begin{eqnarray}
K &=&\frac{1}{2\omega _{\Lambda }t_{\Lambda }}\sum_{l}\int_{\mathbf{r}}\left[
\left\vert \mathbf{D}\phi _{l}\right\vert ^{2}+d^{-2}\left\vert \phi
_{l}-\phi _{l+1}\right\vert ^{2}\right.   \notag \\
&&\left. +\left( 2\varepsilon -b\right) \left\vert \phi _{l}\right\vert ^{2}
\right] \text{,}
\end{eqnarray}%
and a small perturbation
\begin{equation}
W=\frac{1}{2\omega _{\Lambda }t_{\Lambda }}\sum_{l}\int_{\mathbf{r}}\left[
\left( t_{\Lambda }+b-1-2\varepsilon \right) \left\vert \phi _{l}\right\vert
^{2}+\left\vert \phi _{l}\right\vert ^{4}\right] \text{.}  \label{V}
\end{equation}%
Here, the variational parameter $\varepsilon $ (that depends on temperature,
magnetic field and material parameters) has a physical meaning of the
excitation gap in the vortex liquid phase. It is found from minimization of
the variational free energy including the fluctuations on the mesoscopic
scale. The only nontrivial technical difficulty is the summation over Landau
levels in the presence of UV cutoff $\Lambda $. It is shown\cite{suppl} that
to absorb all UV divergences one has to sum over Landau levels till the
"maximal" one $N_{\max }=\Lambda /b-1$. This results in the vortex liquid
gap equation
\begin{eqnarray}
\varepsilon  &=&\frac{t_{\Lambda }+b-1}{2}+\frac{\omega _{\Lambda
}t_{\Lambda }d}{2\pi ^{2}}\int_{k=0}^{2\pi /d}\left\{ \psi \left( g+\Lambda
/b\right) -\psi \left( g\right) \right\} ;  \label{gapeq} \\
g &\equiv &\left( 1-\cos (kd)\right) /\left( d^{2}b\right) +\varepsilon /b%
\text{,}  \notag
\end{eqnarray}%
where $\psi $ is the digamma function. The integration is over the Fourier
harmonics $k$ in the $c$ direction.

The SCFT is widely used in GL model without magnetic field, $b=0$, under the
name of "mean field" and in this case simplifies to
\begin{eqnarray}
\varepsilon  &=&\left( t_{\Lambda }-1\right) /2+\omega _{\Lambda }t_{\Lambda
}\left( h\left( \Lambda +\varepsilon \right) -h\left( \varepsilon \right)
\right) \text{;}  \notag \\
h\left( u\right)  &=&\ln \left( 1+ud^{2}+\sqrt{2ud^{2}+\left( ud^{2}\right)
^{2}}\right) /\pi \text{.}  \label{gap_eq}
\end{eqnarray}%
In this case $\varepsilon $ has a meaning of the "mass" of the field $\phi $
describing the fluctuations in the normal phase. It vanishes at the
"renormalized" transition temperature $T_{c}$ leading to its relation to $%
T_{\Lambda }$
\begin{equation}
T_{\Lambda }^{-1}=T_{c}^{-1}\left( 1-2\omega h\left( \Lambda \right) \right)
\text{.}  \label{Tc}
\end{equation}%
Here the renormalized coupling $\omega =\sqrt{2Gi}\pi /s$, this time
expressed via renormalized Ginzburg number $Gi=2\left( e^{\ast }/c\hbar
\right) ^{3}\kappa ^{4}T_{c}^{2}\gamma ^{2}/H_{c2}$, is used. Expressing $%
T_{\Lambda }$ via $T_{c}$ in Eq.(\ref{gapeq}), the gap equation becomes,
\begin{eqnarray}
\varepsilon  &=&\frac{\omega td}{2\pi ^{2}}\int_{k=0}^{2\pi /d}\left[ \psi
\left( g+\Lambda /b\right) -\psi \left( g\right) \right]   \notag \\
&&-\omega th\left( \Lambda \right) +\left( t+b-1\right) /2\text{,}
\label{renorm_gapew}
\end{eqnarray}%
with $t=T/T_{c}$. Physical quantities are then calculated using numerical
solution of this algebraic equation. For $b,\varepsilon <<\Lambda $ it is
cutoff independent and simplifies:
\begin{equation}
\varepsilon =\left( t+b-1\right) /2-\frac{\omega td}{2\pi ^{2}}%
\int_{k=0}^{2\pi /d}\left[ \psi \left( g\right) +\ln 2\right] \text{.}
\label{crit_gap_eq}
\end{equation}

Magnetization is\cite{suppl},
\begin{gather}
M=\frac{\omega stH_{c2}}{8\pi ^{3}\kappa ^{2}}\int_{k=0}^{2\pi /d}\left[
\left( g+\Lambda /b-1/2\right) \psi \left( g+\Lambda /b\right) \right.
\notag \\
\left. -\left( g-1/2\right) \psi \left( g\right) +\ln \left( \Gamma \left(
g\right) /\Gamma \left( g+\Lambda /b\right) \right) -\Lambda /b\right] \text{%
,}  \label{m}
\end{gather}%
while for $b,\varepsilon <<\Lambda $ it simplifies to
\begin{equation}
M=\frac{\omega stH_{c2}}{8\pi ^{3}\kappa ^{2}}\int_{k=0}^{2\pi /d}\left[ \ln
\frac{\Gamma \left( g\right) }{\sqrt{2\pi }}+g-\left( g-\frac{1}{2}\right)
\psi \left( g\right) \right] \text{.}  \label{m_appr}
\end{equation}%
In certain portions of the magnetic phase diagrams the strong inequalities $%
b,\varepsilon <<\Lambda $ are not obeyed, while SCFT is
still valid, so we have used the formula Eq.(\ref{m}), with weak
(logarithmic) cutoff dependence instead of the cutoff independent
renormalized formula.

\textit{Comparing with experiments and GFT.} Recent accurate magnetization
data \cite{LuLi10} on magnetization of three major families of HTSC
materials, including underdoped $La_{2-x}Sr_{x}CuO_{4}$ for $x=0.09$,
optimally doped $Bi_{2}Sr_{2}CaCu_{2}O_{8+\delta }$, and optimally doped $%
YBa_{2}Cu_{3}O_{7}$, are fitted in Fig. 2a, 2b, 2c respectively. Measured
magnetization curves of $LSCO$ and $YBCO$ in the $0-14T$ field range and $%
BSCCO$ at $0-40T$ show distinct features above and below $T_{c}$, thus
allowing meaningful fitting. The conditions $b,\varepsilon <<\Lambda $ are
obeyed provided\ temperature does not deviate too far from $T_{c}$ and
magnetic field is small compared to $H_{c2}$. Several temperatures within $%
10\%$ of $T_{c}$ were used to determine three fitting parameters, $H_{c2}$,
anisotropy $\gamma $, and $\kappa ^{2}/s$, using simplified formulas Eqs.(%
\ref{crit_gap_eq}, \ref{m_appr}). The interlayer distances $d^{\prime }$
were taken from \cite{Poole2007}. Near $T_c$, the correlation length is large, therefore we take $s=d$, as the maximum value of $s$. The results for each material are given in Table I.

\begin{figure}[tbp]
\begin{center}
\includegraphics[width=7cm]{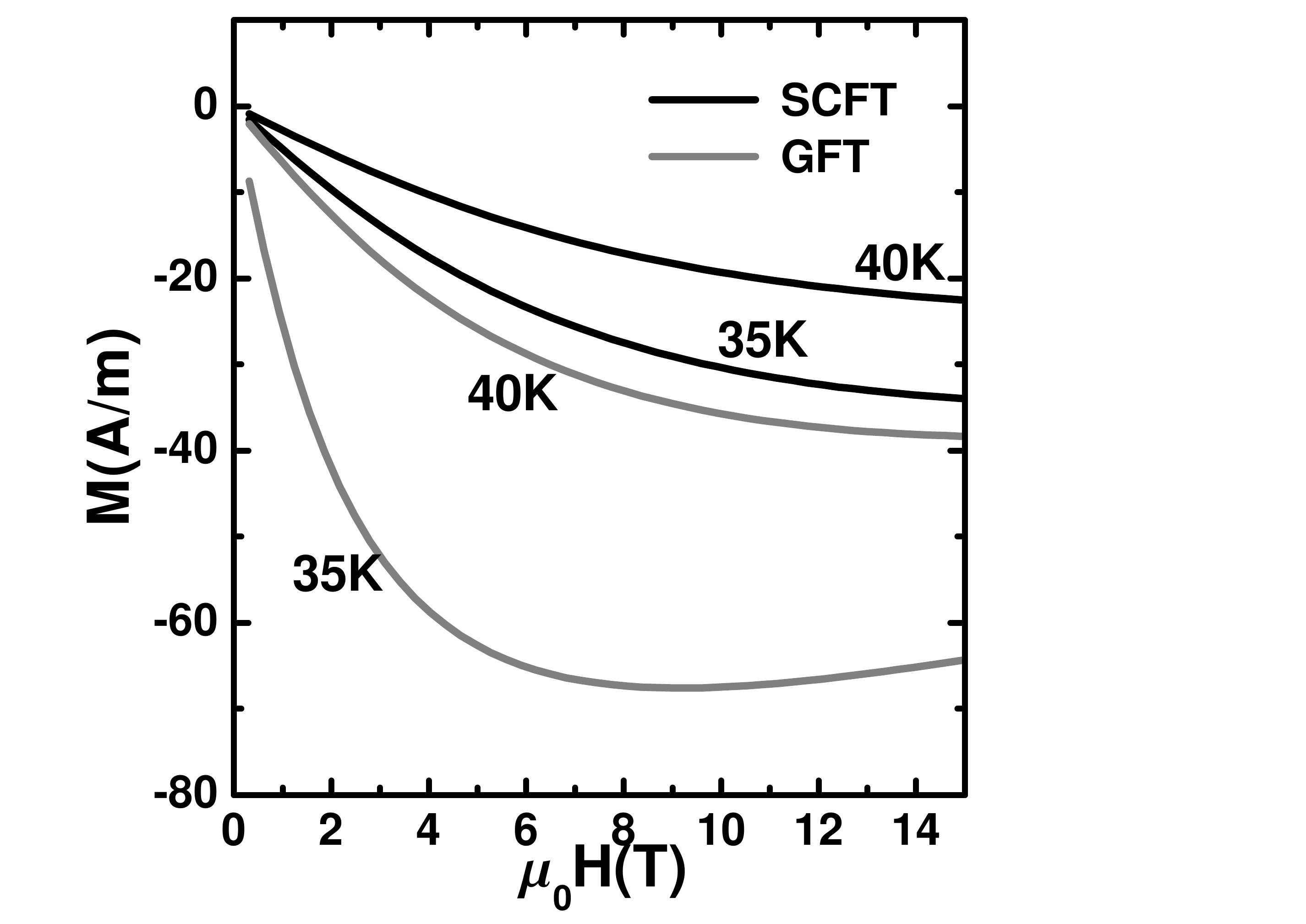}
\end{center}
\par
\vspace{-0.5cm}
\caption{The comparison between the fluctuation magnetization of $LSCO$
calculated using the self consistent fluctuation theory (SCFT) \textit{vs} the
perturbative gaussian fluctuation (GFT) one. }
\end{figure}

For the rest of the data (higher temperature and higher magnetic field) the
theoretical curves shown in Fig. 2. were logarithmically dependent on cutoff
and therefore the full formulas, Eqs.(\ref{gapeq}, \ref{m}), were utilized.
The two additional parameters, namely mean field critical temperature $%
T_{\Lambda }$ and $\Lambda $ are constrained via Eq.(\ref{Tc}) (with
experimentally measured $T_{c}$ also listed in Table I). The values of $%
T_{\Lambda }$ and $\Lambda $ in units of $\hbar e^{\ast }H_{c2}/(m_{a}c)$
are given in Table I.

To demonstrate the importance of nonperturbative effects the SCFT
magnetization, Eq.(\ref{m}) is compared with GFT within the 2D layered
superconductors model \cite{Carballeira00} in Fig. 3. One observes that The
SCFT magnitude is much smaller than the GFT one. One of the reasons is that
the vortex liquid gap $\varepsilon $ is larger than the reduced temperature
(perturbative gap) $\left( t_{\Lambda }+b-1\right) /2$.

The data of ref. \cite{LuLi10} in the region of smaller fields exhibit the
so called "intersection point" of the magnetization curves plotted as
function of temperature. Our magnetization curves (underdoped $LSCO$ is
shown in Fig. 4 as an example) demonstrate the intersection point in this
region for all three materials. The intersection points were measured in
many high $T_{c}$ cuprate \cite{Salem} and explained within the "lowest
Landau level" approximation \cite{Lin05} valid for $\varepsilon <<b$. It
turns out that an addition requirement for the intersection point is $%
\varepsilon d^{2}>>1$. Our results demonstrate that beyond this
approximation the intersection point disappears.

\begin{figure}[tbp]
\begin{center}
\includegraphics[width=7cm]{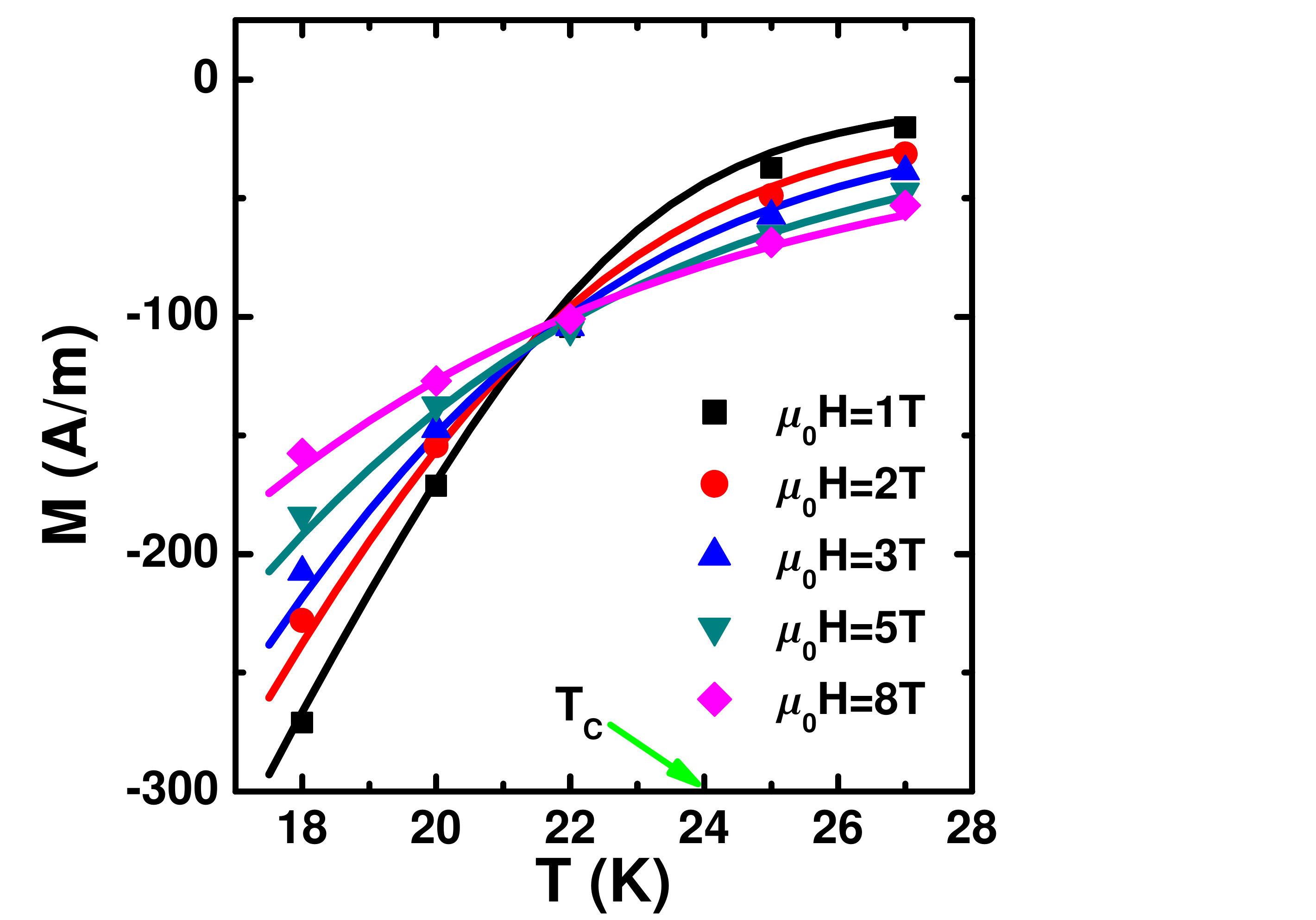}
\end{center}
\par
\vspace{-0.5cm}
\caption{Fit of magnetization \protect\cite{LuLi10} in the region of the
intersection point (fields lower than those shown in Fig. 2a) in LSCO using
the same fitting parameters (given in Table I).}
\end{figure}

\textit{Conclusions.} We have investigated the fluctuation diamagnetism of
HTSC using a self consistent nonperturbative method beyond gaussian
fluctuations term within Lawrence - Doniach GL model. The comparison with
recent accurate experiments near $T_{c}$ demonstrate that the effect of
quartic terms should to be included due to strong fluctuations. The theory
describes well wide class of materials from relatively low anisotropy
optimally doped $YBCO$ to highly anisotropic underdoped $LSCO$ and optimally
doped $BSCCO$ at temperatures both below and above $T_{c}$. No input from
the microscopic "pseudogap" physics is needed to describe the magnetization
data. Dynamical effects like Nernst effect, electrical and thermal
conductivity can be in principle approached within the similar SCFT
generalized to a time dependent variants of the GL model.

\begin{acknowledgments}
The work of DL and XJ is supported by National Natural
Science Foundation of China (No. 11274018), BR is supported by NSC of R.O.C.
(No. 8907384-98N097).
\end{acknowledgments}


\end{document}